\documentclass[twocolumn,nofootinbib,prd,floatfix,preprintnumbers,amsmath,amssymb,groupedaddress,superscriptaddress]{revtex4}
\usepackage{dsfont}
\usepackage{epsfig}
\usepackage{slashed}
\usepackage{bbold}
\usepackage{feynmp}
\usepackage{booktabs}
\usepackage{array}
\usepackage{multirow}

\begin{document}
\setlength{\unitlength}{1mm}
\title{Flavourful Production at Hadron Colliders}

\author{Gian Francesco Giudice}
\affiliation{CERN PH-TH, Geneva 23, 1211 Switzerland.}

\author{Ben Gripaios}
\affiliation{CERN PH-TH, Geneva 23, 1211 Switzerland.}

\author{Raman Sundrum}
\affiliation{Department of Physics, University of Maryland, College Park, MD 20742, USA.}
\preprint{CERN-PH-TH/2011-110}
\preprint{UMD-PP-11-006}
\begin{abstract}
We ask what new states may lie at or below the TeV scale, with sizable flavour-dependent
couplings to light quarks, putting them within reach of hadron colliders via resonant production, or in association with Standard Model states. 
In particular, we focus on
 the compatibility of such states with stringent flavour-changing neutral current and electric-dipole moment constraints. 
We argue that the broadest and most theoretically plausible flavour structure of the new couplings is that they are hierarchical, as are Standard Model Yukawa couplings, although the hierarchical pattern may well be different. 
 We point out that, without the need for any more elaborate or restrictive structure,
new scalars with ``diquark'' couplings to standard quarks are particularly immune to 
existing constraints, and that such scalars may arise within a variety of theoretical paradigms. In particular, there can be substantial couplings to a pair of light quarks or to one light and one heavy quark. For example, the latter possibility may provide a flavour-safe interpretation of the asymmetry in top quark production observed at the Tevatron.
We thereby motivate searches for diquark scalars at the Tevatron and LHC,  and argue that their discovery represents one of our best chances for new insight into the Flavour Puzzle of the Standard Model.
\end{abstract}
\maketitle
\section{Introduction\label{sec:intro}}
For some decades now, phenomenological research into physics beyond the Standard Model (SM) has been driven by the question of `What new physics should we see?' That is to say, the SM leaves unresolved several deep mysteries,  
such as the electroweak hierarchy problem, the flavour puzzle, the identity of dark matter, the origin of the matter-antimatter asymmetry,  \&c, 
and new models that address these issues inevitably
predict new degrees of freedom. Diverse and intense experimental efforts have been 
directed towards their discovery. Thus far however, although there are some intriguing anomalies, there has been no decisive experimental break with the SM.\footnote{Neutrino oscillations are a notable exception. 
While these data provide valuable new clues on the flavour puzzle, it can be accommodated
by only a modest extension of SM structure.}
 Indeed, the myriad null searches (both direct and virtual)  now very tightly constrain the form of any new physics near the TeV scale. 
Our dominant inspiration, the hierarchy problem, has led to beautiful new ideas and the vision of a rich new spectrum, but proposed resolutions only remain viable in small regions of their natural parameter space.  

Happily, the game is far from over. Analysis of the large Tevatron data set is ongoing, and the LHC has begun the most thorough exploration yet of the TeV scale. Hopes are high for the discovery of new physics. 
It may be that an unforeseen, but fully satisfying, resolution to the hierarchy problem emerges directly from the data. 
Or it  may be that standard considerations of the hierarchy problem and  fine-tuning provide only a very crude guide as to where new physics should appear. In that case, one of the existing paradigms may yet be discovered, with parameters that seem to us ``tuned'' at the {\em per cent} level.  But life could be more complicated. There may well be a rich new threshold at which the lion's share of the hierarchy problem is resolved, but it may lie above LHC reach, say at 10 TeV, or 100 TeV, with $10^{-4}$ tuning by effective field theory measure.
 In that case, even the LHC  will catch only those associated states
that happen to be exceptionally light,  an opening move but not the end-game. A rough parallel can be found in the SM itself:
among the states that feel electroweak breaking
 we first discovered the electron because it was one of the lightest, even though it was not a ``major player''   of the likes of the $W, Z$, top or Higgs. If the LHC, at least in its early stages, is sensitive only to a light vestige of a major mechanism at higher energies, the question shifts from `What new physics {\em should} we see?' to the more humble `What new physics {\em could} we see?' 
More precisely: what new physics (however dimly perceived its ultimate ``purpose'') is consistent, without too much theoretical artifice, with the many existing constraints?

There are three reasons for trying to develop a systematic approach to this question.
Firstly, the constraints on TeV physics are so strong that, subject to some plausible  assumptions, the litany of new physics possibilities may be short enough to catalogue.
Secondly, the complexities of hadron machines are such that we are unlikely to discover anything without expressly looking for it.  
Thirdly, such a catalogue of allowed, albeit ``unmotivated'', possibilities is of theoretical use, in highlighting new paradigms, mechanisms and modules which guarantee consistency with existing constraints. It is to this general programme that we will contribute in this paper.

Let us first establish some ground rules, whose purpose is to balance the observability of TeV-scale new physics against its plausible connection to some deeper theoretical ``plot".
While scenarios such as weak scale supersymmetry predict a doubling or more of the SM spectrum, we  contemplate here a much more modest number of new particles with sub-TeV masses, with perhaps a richer but heavier spectrum out of range.
We will also focus on particles that can be readily and directly produced at colliders, with appreciable coupling to the SM, in particular to SM gauge bosons and/or
to SM light fermions. This balances experimental observability with 
our best guess that a light vestige of some (generally heavy) solution to the SM hierarchy problem should couple directly to the SM.\footnote{This  is in contrast to, say, a ``hidden valley'' structure \cite{Strassler:2006im} in which there are light states that do not directly couple to the SM, but are produced via a heavy ``bridge'' particle, which does couple to the SM. In terms of observability, the production cross-sections are low, but off-set by the spectacular nature of the events.}

 At the technical level, the requirement that the new particles  couple substantially to the SM implies that these couplings are {\it renormalizable} in form.  
 This is because non-renormalizable couplings rapidly weaken below the scale of their UV completion, and we are contemplating that any such higher physics scale is at least above LHC reach. Non-renormalizable interactions may however play an important role in the decays of a new particle, especially if its renormalizable interactions alone would leave it stable.  Renormalizability restricts the new particles to spin $\leq 1$.
New particles with SM gauge quantum numbers then certainly represent a very plausible possibility, and also a very safe option from the point of view of existing constraints. 
Gauge interactions are famously ``flavour-blind'' and so this kind of new physics can readily evade stringent FCNC constraints, while still allowing pair-production (or even resonant production \cite{Kilic:2008pm,Kilic:2009mi}) of  new particles. Very weak or non-renormalizable couplings to the SM may subsequently mediate the decay of these particles, with only mild constraints from flavour physics. Another possibility is that the new particles are themselves spin-1 gauge bosons under which some of the SM is charged.
Again, the choice of gauge couplings can naturally be sufficiently flavour-blind, but 
 constraints from electroweak precision tests
or searches for jet or lepton excesses can still be challenging. 

Gauge interactions are the only renormalizable interactions until one introduces scalars.
 Minimally, this could be just the SM Higgs boson, to which new particles can couple.
But one might also have a new scalar. Beyond SM gauge interactions, a new scalar can also have
 sizable Yukawa couplings to a pair of SM fermions, or to a SM fermion and a new fermion.
 Such Yukawa couplings might well provide a new window on flavour physics. 
Indeed, one might worry that the vast array of past flavour tests already
 severely constrains this possibility unless there is a very special mechanism in place.

Let us survey the possibilities for such a scalar, 
 $\phi$. Denoting SM quarks by $q$, SM leptons by $\ell$, and possible new fermions by 
$\chi$, we see that the list of $\phi$ Yukawa couplings with at least one SM fermion is given by (i) $\phi \bar{q} q, \phi \bar{\ell} \ell$, (ii) $\phi \bar{q} \ell, 
\phi \bar{q} \bar{\ell}$, (iii) $\phi q \chi, \phi \ell \chi,$  (iv) $ \phi \ell \ell$, (v) $\phi qq$. The $\phi$ of category (i) necessarily has the same electroweak quantum numbers of the SM Higgs multiplet, and is either colour octet \cite{Manohar:2006ga} or colour singlet. It is well understood that such generalized Higgses can all too readily mediate excessive FCNCs if they lie below a TeV, unless their couplings are very carefully chosen. If there is a second colour-singlet Higgs doublet, the Glashow-Weinberg rule \cite{Glashow:1976nt} for designating one as an up-type Higgs and the other as down-type, as enforced for example by supersymmetry, ensures the absence of tree-level Higgs-mediated FCNCs. More generally, a much tighter Yukawa-coupling {\em ansatz}, such as
Minimal Flavour Violation (MFV) \cite{D'Ambrosio:2002ex}, is required to suppress tree-level FCNCs. It requires quite specialized structure in the UV to arrange for it. Even a small deviation from the MFV structure can be deadly, leading to excessive FCNCs.
Category (ii) $\phi$s are 
``leptoquarks''. Such couplings can be relevant for decays at hadron colliders \cite{Gripaios:2009dq,Davidson:2010uu,Gripaios:2010hv, Davidson:2011zn}, but unlikely to dominate production. Sfermions of supersymmetry 
are examples of category (iii), with $\chi$ being a gaugino or Higgsino. Of course, this case is famously dangerous for FCNCs, as encapsulated in the supersymmetric flavour problem. Again, special patterns such as gauge-mediated supersymmetry breaking 
can evade this generic concern.\footnote{Non-flavour blind, flavourful alternatives were discussed in \cite{Nomura:2007ap}.}  Category (iv) $\phi$s are ``dileptons'', which are less relevant to hadronic collider production. Finally, category (v) $\phi$s are 
``diquarks''. Such couplings, if strong enough, can play an important role in new physics production at hadronic colliders  \cite{Ma:1998pi,Bauer:2009cc}. 
Indeed, they have already been suggested by several authors \cite{Shu:2009xf,Dorsner:2009mq,Dorsner:2010cu,Gresham:2011dg,Patel:2011eh,Grinstein:2011yv,Ligeti:2011vt} as an explanation for the $t\overline{t}$ forward-backward asymmetry.
They are the focus of this paper.

Na\"{\i}vely, such diquark couplings, $\phi qq$, appear to pose similar FCNC concerns as Higgs-like couplings, $\phi \bar{q} q$ \cite{Ma:1998pi}, evaded only by the same level of specialization of couplings, such as MFV \cite{Arnold:2009ay}. Remarkably, this is {\it not} necessarily the case. For example, consider a colour-triplet scalar that couples to right-handed up-type quarks, $\phi u_R^i u_R^j$, where $u^{i=1,2,3} \equiv u, c, t$. 
QCD gauge invariance implies colour antisymmetry, which in turn implies flavour antisymmetry of the quarks. 
Now, antisymmetry implies that any flavour-changing diagram must involve all three generations of quarks. Indeed, with only two quark generations, the diquark
coupling is proportional to $\epsilon_{ij}$, which is an invariant tensor of the $SU(2)$ flavour symmetry acting on $u_R^i$. This has various consequences. For example, tree-level $\phi$ exchange cannot mediate $\Delta F=2$ FCNCs, since only two generations would be involved. 
Even at one-loop, the contributions to neutral $D$-meson mixing must proceed via the third generation. Thus, they do not involve the $\phi u_R c_R$ coupling, which consequently may be of order unity, even for a diquark mass of a few hundred GeV, resulting in a huge production cross section at the LHC. As we shall see, there are also remarkable suppressions of flavour-diagonal, electric dipole moments (EDMs), in that diagrams involving only quarks and diquarks do not contribute at {\em any} loop order. 

Such considerations make a much broader and more plausible flavour structure of diquark couplings possible.
Indeed our purpose here is to point out that scalars with substantial diquark couplings represent a unique combination of experimental and theoretical
opportunity at hadron colliders such as the LHC and Tevatron. In addition to their QCD couplings, the diquark couplings can be strong enough to play an important role in  new physics production and decay. The fact that these couplings can tolerate a variety of patterns without already being ruled out by flavour and other precision data means that experiments can teach us something about flavour structure that we do not already know. (By contrast, in the MFV {\em ansatz}, precisely the flavour structure of SM Yukawa couplings is replicated in new physics.) Finally, such ``diquark" scalars can very plausibly represent the low end of a new spectrum which addresses the ``big'' hierarchy problem.
For example, the scalars might be (i) pseudo-Goldstone bosons of a new strong dynamics which makes a composite Higgs boson, (ii) squark remnants of 
supersymmetric physics, with $R$-parity violating couplings to quarks, (iii) a coloured partner of the SM Higgs boson in some (orbifold) unification schemes \cite{Cheung:2002uz}. Or they may be a light vestige of something unanticipated. 

It is certainly not the case that {\em any} flavour structure of diquark couplings is consistent with FCNC constraints.  Instead, we will be guided very  broadly by what we already see in the  flavour structure of SM quark Yukawa couplings. The hierarchical pattern of quark masses and mixing angles strongly suggests that there is a particular electroweak gauge basis in which  the Yukawa matrices have a hierarchical structure of matrix elements, such that the first, second, and third generations have an increasing degree of connectedness to the SM Higgs boson. While the SM itself offers no explanation of this fact, it can be understood in a variety of UV theories in which these three quark generations are distinguished.  For example, in Froggatt-Nielsen theory \cite{Froggatt:1978nt}, the 
hierarchical structure arises from assigning the
 three generations distinct charges under a Higgsed $U(1)_{FN}$ gauge theory. In higher-dimensional theories, hierarchy arises when the three generations have distinct extra-dimensional wavefunctions with varying overlaps with the Higgs boson \cite{ArkaniHamed:1999dc}.
In strongly-coupled (or AdS/CFT-related warped) models, the three generations are distinguished by scaling dimensions \cite{Kaplan:1991dc,Grossman:1999ra,Gherghetta:2000qt,Nelson:2000sn,Huber:2000ie,Huber:2003tu}.

The minimal structure we will impose on diquark couplings is that
they and the SM Yukawa matrices are hierarchical in the 
 same electroweak gauge basis. They need not, however, exhibit the {\em same} hierarchical structure. This seems to us the most plausible and broadly phrased approach to how new physics flavour structure might appear at collider energies. Note that this assumption in no way saves all types of new scalars from danger.
For example, new Higgs bosons with hierarchical couplings that approximately mimic SM Yukawa couplings would satisfy our criteria, but would give tree-level FCNCs, typically far in excess of experiment. Only a much more restrictive {\em ansatz}, such as MFV, can avoid this.  
More generally, even if we started with an extreme hierarchy of couplings for a new Higgs-like scalar in the gauge basis, with only a single, non-vanishing, diagonal entry in the coupling matrix,  the rotation required to go to the quark mass basis would already lead to tree-level FCNCs, again typically excessive. But we will show that scalars with diquark couplings are much safer. So safe in fact that, at least for antisymmetrically-coupled diquarks, any one of the couplings could be the largest. Thus, the largest coupling may involve quarks of the second and third, the first and second, or the first and third generations and each gives rise to a distinct phenomenology. We call the resulting hierarchies the normal, inverted, and perverted hierarchies, respectively.
The perverted hierarchy, in particular, can be used to explain the asymmetry in $t\overline{t}$ production in a way that is consistent with constraints from $D$-meson mixing and single top production. 

The remainder of this paper is organized as follows. In \S \ref{sec:para}, we will further frame our flavour philosophy, and illustrate how it naturally fits with popular UV approaches to flavour structure. In \S \ref{sec:clas}, we classify diquarks by their SM gauge charges, 
tree-level FCNCs, and proton stability. In \S \ref{sec:loop}, we consider loop-level effects. In \S \ref{sec:skew}, we restrict ourselves further to the two states with (automatically) antisymmetric flavour couplings and determine existing bounds on the masses and couplings coming from one-loop processes and from tree-level, flavour-violating decays.
We discuss flavour-diagonal processes and electroweak precision tests in \S \ref{sec:diag} and outline strategies for collider searches for the various paradigms in \S \ref{sec:coll}.
\section{Flavour philosophy\label{sec:para}}
We begin our journey with the Yukawa couplings of the SM, which generate the observed masses and mixings of quarks and provide the origin of flavour in the SM. As is well-known, these masses and mixings are not anarchical, but rather possess a curious pattern of hierarchies and degeneracies. 

One {\em ansatz} for the Yukawa couplings that leads to a good description of the observed masses and mixings takes the form (see {\em e.\ g.\ }\cite{Grossman:1999ra,Gherghetta:2000qt,Nelson:2000sn,Huber:2000ie,Kaplan:2001ga,Huber:2003tu,Davidson:2007si})
\begin{gather} \label{basis}
\mathcal{L} = -\Sigma_{i,j} \left( y^u_{ij} \epsilon^q_i \epsilon^u_j \overline{q}_L^i H u_R^j + y^d_{ij} \epsilon^q_i \epsilon^d_j \overline{q}_L^i H^c d_R^j  \right) + \mathrm{h. c.} ,
\end{gather}
where the $\epsilon^{q,u,d}_i \leq 1$ are hierarchical between the different generations (though not necessarily between the different quark multiplets) and the coefficients $y_{ij}$ are of order one. We will refer to the Yukawa structure of eq.~(\ref{basis}) as the {\em Chiral Hierarchy}.

Passing to the mass basis, we find that both $\epsilon^q_3$ and  $\epsilon^u_3$ should be of order one in order to reproduce the large top quark mass, while the CKM matrix is given by
\begin{gather} \label{mixings}
V_{CKM} \sim \begin{pmatrix} 1& \frac{\epsilon^{q}_{1}}{\epsilon^{q}_{2}}&  \frac{\epsilon^{q}_{1}}{\epsilon^{q}_{3}} \\ -\frac{\epsilon^{q}_{1}}{\epsilon^{q}_{2}}& 1&  \frac{\epsilon^{q}_{2}}{\epsilon^{q}_{3}}\\  - \frac{\epsilon^{q}_{1}}{\epsilon^{q}_{3}} &-  \frac{\epsilon^{q}_{2}}{\epsilon^{q}_{3}} & 1 \end{pmatrix}.
\end{gather}
The three measured mixings, $V_{cb}$, $V_{us}$, and $V_{ub}$, then fix  $\epsilon^q_2 \sim \lambda_{\mathrm{CKM}}^2$ and  $\epsilon^q_1\sim \lambda_{\mathrm{CKM}}^3 $, where $\lambda_{\mathrm{CKM}} \sim 0.23$ is the Cabibbo angle, together with the successful ``prediction" $V_{ub} \sim V_{cb} V_{us}$. We may then solve for the remaining five $\epsilon$ parameters in terms of the remaining five quark masses. In all, we have
\begin{align} \label{masses}
\epsilon^{d}_{3} &\sim  \frac{m_b}{m_t}, \\
 \epsilon^{d}_{2} &\sim  \frac{m_s}{m_t  V_{cb}} ,  \; \;  \epsilon^{u}_{2} \sim  \frac{m_c}{m_t V_{cb}} , \\
 \epsilon^{d}_{1} &\sim  \frac{m_d}{m_t V_{us} V_{cb}}  ,  \; \;  \epsilon^{u}_{1} \sim  \frac{m_u}{m_t V_{us} V_{cb}} .
\end{align}
In the SM, the only physical parameters arising from the rotation are contained in the quark masses and the CKM matrix; when we add new physics states coupled to quarks, other parts of the quark rotations also become physical. In this way, new physics could also be important for further elucidating the flavour structure within the SM: the presence of new physics is necessary to render the right handed rotations measurable, but these rotations are part of the same underlying structure that generates the SM flavour parameters. 
In particular, Chiral Hierarchy (\ref{basis}) predicts right-handed rotations which take the same form as (\ref{mixings}), but with $\epsilon^{q}_i \rightarrow \epsilon^{u,d}_i$. Thus,
\begin{align} \label{rmixings}
&V^{u_R}_{12} \sim \frac{m_u}{m_c V_{us}}\sim 0.01, \; \; V^{u_R}_{23} \sim \frac{m_c}{m_t V_{cb}}\sim 0.09, \nonumber \\
&V^{d_R}_{12} \sim \frac{m_d}{m_s V_{us}}\sim 0.2, \; \; V^{d_R}_{23} \sim \frac{m_s}{m_b V_{cb}}\sim 0.5, \nonumber \\
&V^{u_R,d_R}_{13} \sim V^{u_R,d_R}_{12} V^{u_R,d_R}_{23},
\end{align}
where in making the estimates we have taken central values for the quark masses at the TeV scale.
While $V^{d_R}_{12}$ is parametrically equal to the Cabibbo angle and $V^{d_R}_{23}$ is larger than $V_{cb}$, $V^{u_R}_{12}$ and $V^{u_R}_{13}$ are much smaller than their CKM counterparts. This asymmetry between the left- and right-handed rotations arises because the Yukawa couplings in Chiral Hierarchy are not symmetric matrices. We shall later see that the suppression of mixings may lead to an important suppression of flavour-changing processes from new physics and will be crucial for explaining the $t\overline{t}$ asymmetry in the context of our flavour paradigm.

While Chiral Hierarchy gives a good fit to the measured hierarchies in masses and mixings, it is unnecessarily restrictive for our needs. In particular, the suppression of right {\em versus} left mixings just described need not be a general feature of hierarchical Yukawa couplings. As a counter-example, in a GUT model with all quarks living in the same multiplet, one might expect the Yukawa matrices to be symmetric or 
antisymmetric and thus the right rotation matrix to be equal to the complex conjugate of the left rotation matrix. As a result, if the rotations in the up and down sectors were comparable, we would have
\begin{align} \label{ckmmixings}
&V^{u_R,d_R}_{12} \sim \lambda_{\mathrm{CKM}}\sim 0.23, \; \; V^{u_R,d_R}_{23} \sim \lambda_{\mathrm{CKM}}^2 \sim 0.05, \nonumber \\
&V^{u_R,d_R}_{13} \sim V^{u_R,d_R}_{12} V^{u_R,d_R}_{23}.
\end{align}
So, in order to be as general as possible, we shall not impose Chiral Hierarchy in what follows. Rather, we shall assume that there exists some pattern of hierarchies in the SM Yukawa couplings and that the resulting mixings in the right handed sector lie somewhere between the values in (\ref{rmixings}), which correspond to Chiral Hierarchy, and the values in (\ref{ckmmixings}), which we refer to as {\em CKM-like mixing}.

Now let us move on to new physics. Our flavour philosophy will be that any new physics should also possess some hierarchical structure in its couplings, {\em in the gauge basis}. From the IR viewpoint, this seems not only reasonable (since the SM Yukawa couplings already possess such a structure), but also necessary (in that anarchic, sizable new couplings would be irreconcilable with flavour physics constraints).

In contrast with the previous literature,\footnote{Ref. \cite{BarShalom:2007pw} differs in considering extra Higgs scalars with sizable couplings involving the light generations. However, since other couplings are set to zero in the mass basis, this does not fall within the scope of our flavour philosophy.} we shall not insist that the hierarchical structure in the new couplings is the same as the structure in the Higgs Yukawa couplings. Rather, we shall take the view that any structure is acceptable, especially given our ignorance about the UV dynamics that generates flavour.

Before continuing, we should address one objection that may irk the reader: why, from the UV viewpoint, should the couplings of both the Higgs and extra scalars to fermions be hierarchical in the same basis? If they are not, then the rotation from one basis to the other would restore anarchy in one or other of the coupling matrices, with disastrous consequences for flavour physics.

We claim that having both coupling matrices hierarchical in the same gauge basis is quite plausible, if one takes the viewpoint that the UV theory of flavour makes a strong distinction between the different fermion generations, as occurs in the examples mentioned in \S \ref{sec:intro}. In an extra-dimensional model, for example, 
a hierarchical flavour structure that arises because of small wavefunction overlaps will appear in the couplings of all scalars, in the same basis.

Another objection to our philosophy might be that, although a hierarchical structure in the Higgs and new scalar couplings is plausible, different hierarchical structures are not. This does not seem to hold water to us either, since different scalars may interact in very different ways with the SM fermions. 
Indeed, the examples in \S \ref{sec:intro} can easily generate different hierarchical structures for different scalars: in the extra-dimensional case, for example, it suffices to localize the scalars in different places; in the Froggatt-Nielsen case, it suffices to assign different flavon charges to different scalars. 

To be explicit, let us sketch how each of these two examples could give rise to the normal, inverted, or perverted hierarchies which we introduced above for antisymmetrically-coupled diquarks, as follows. In the extra-dimensional example, imagine that both scalars and fermions have wavefuctions that are exponentially localized (in some metric), but that the profiles of fermions are broader than those of the scalars. The Yukawa couplings then take exactly the form in (\ref{basis}), where the $\epsilon$ parameters are given by the values of the respective fermion wavefunctions evaluated at the location of the scalar. Thus, the third generation is simply the fermion that lives closest to the Higgs, while the first generation is furthest away and the second generation is somewhere in between.  Then, to obtain the normal hierarchy, it suffices to localize the new scalar closer to the second and third generations than to the first. The other hierarchies can be obtained by moving the location of the new scalar around in an obvious way. 

In the Froggatt-Nielsen example, suppose that the flavon symmetry is a $U(1)_{FN}$, under which the Higgs is neutral and 
$q_L^i, u_R^{ci}$ and $d_R^{ci}$ have positive semi-definite charges $Q^{q,u,d}_i$. 
The $\epsilon$ parameters are then given by $\epsilon_i = \delta^{Q^{q,u,d}_i}$, where $\delta$ is the ratio of the flavon VEV to the cut-off. To get the large top mass requires $Q^{q,u}_3 = 0$. One may then obtain any one of the three hierarchies by an appropriate choice of a negative charge, $Q$, for the new scalar, since the coupling to quarks of generations $i$ and $j$ will be suppressed by $\delta^{|Q+Q_i+Q_j|}$.
For example, in the simplest case in which $q_L^i, u_R^{ci}$ and $d_R^{ci}$ have common charge $Q_i$, the choice
$Q + Q_1 + Q_3 = 0$, corresponds to the perverted hierarchy with subdominant couplings (in the gauge basis) given by $\lambda_{12} = \sqrt{\frac{m_2}{m_3}}$ and   $\lambda_{23} = \sqrt{\frac{m_1}{m_3}}$.\footnote{These examples show that, whenever the UV physics has a non-trivial flavour structure, an effective theory-approach where only SM fields are retained is, in general, unable to determine correctly the coefficients of non-renormalizable, flavour-violating operators. For example, in the case of a Froggatt-Nielsen model, once we integrate out the new physics, the coefficients of the effective operators involving SM fields will be typically suppressed by powers of $S^\dagger S / \Lambda^2$, where $S$ is the flavon and $\Lambda$ is the cut-off. Such suppression factors cannot be determined just by counting the Froggatt-Nielsen charges of the SM fields alone. Our extra-dimensional example shows that the same is true also in that case, because new-physics interactions and SM Higgs Yukawa couplings can be sensitive to different parts of the quark profiles. A simple counting of the suppression factors in the coefficients of the non-renormalizable, flavour-violating operators, based on powers of quark wave-functions, would give an incorrect result. The specific features of the UV flavour theory are essential for determining the pattern of flavour violation in the quark sector and these cannot be captured by an effective-theory analysis.  A similar result was obtained, in the supersymmetric context, in \cite{Dudas:2010yh}.}

Before closing this section, we stress that we do not wish to restrict ourselves to an explicit flavour model. Indeed, we mention them here only to motivate our generalized flavour paradigm. 
\section{Classification, proton decay, and tree-level FCNC\lowercase{s} \label{sec:clas}}
If we consider scalars coupled to a pair of quarks, via a Yukawa coupling, then there are eight possibilities, distinguished by the representation under the SM gauge group and listed in Table \ref{diquark2}.\footnote{Vector diquarks, either colour triplets or sextets, could couple at dimension four to $\overline{q}u^c$ or $\overline{q}d^c$, but will be ruled out by tree-level FCNC constraints as {\em per} the discussion for scalar states.}

An important observation is that diquark couplings to $q_Lq_L$, $u_R u_R$ and $d_R d_R$ imply symmetry or antisymmetry of the corresponding Yukawa matrices in flavour space in the gauge basis. 
We note that the symmetry properties under interchange of flavour indices are preserved under flavour rotations in $SU(3)_{q}\times SU(3)_{u} \times SU(3)_{d}$. This further implies that the pure up- or down-type couplings (to $u_R u_R$, $d_R d_R$, and $u_L u_L,d_L d_L \subset q_L q_L$) retain their symmetry properties in the quark mass basis. (The mixed up/down-type $u_L d_L \subset q_L q_L$ couplings do not, because distinct rotations on $u_L$ and $d_L$ are required to go to the quark mass basis).
\begin{center}
\begingroup
\begin{table*}[ht]
\begin{ruledtabular}
\begin{tabular}{c  c  c  c  c  c  c c }
Name &  $SU(3)_c $ & $SU(2)_L $   & $U(1)_Y $ & QQ Coupling & LQ Coupling & Tree-level $\Delta F = 2$?  \\
\hline
I&   $\overline{6}$ & 3&$-\frac{1}{3}$ &  $(q_L q_L)$   & - & Yes\\
II&  3 & 3& $-\frac{1}{3}$& $[q_L q_L]$  & $q_L l_L$ & No\\
III&  $\overline{6}$ &1 &$-\frac{1}{3}$ & $[q_L q_L]$, $u_R d_R$  & -&  No\\
IV&  3 & 1 & $-\frac{1}{3}$ & $(q_L q_L)$,$u_R d_R$  & $q_L l_L,u_R e_R$ &  No\\
V&  $\overline{6}$ & 1& $-\frac{4}{3}$& $(u_R u_R)$ & - &  Yes\\
VI&  $3$ &1 & $-\frac{4}{3}$& $[u_R u_R]$ & $d_R e_R$ &  No \\
VII&   $\overline{6}$ & 1&$ \frac{2}{3}$ & $(d_R d_R)$ & - &   Yes\\
VIII&  $3$ &1 & $\frac{2}{3}$ & $[d_R d_R]$   & - & No\\
\end{tabular}
\end{ruledtabular}
\caption{Scalar diquarks and their couplings. The parentheses in the `QQ Coupling' column indicate whether the relevant coupling is symmetric () or antisymmetric [] in flavour indices. \label{diquark2}}
\end{table*}
\endgroup
\end{center}

\subsection{Proton decay}
We now open a parenthesis, to make a short discussion of proton decay, mediated by the new scalars. As indicated in Table \ref{diquark2}, three of the eight possible diquark states (namely II, IV and VI) can also have a dimension-four, Yukawa-type coupling to a lepton and a quark \cite{Ma:1998pi,Arnold:2009ay}. The presence of both the diquark and leptoquark coupling would permit decay of the proton via a dimension-six operator mediated by tree-level exchange of the scalar, implying a very large bound on its mass. This can be easily avoided by declaring that some additional global symmetry stabilizes the proton and forbids the leptoquark coupling. The simplest such symmetry is a baryon parity, under which only the quarks are odd.

For the remaining five diquark states, baryon number is an accidental symmetry of the renormalizable lagrangian, such that we may remain agnostic about its status in the UV \cite{Ma:1998pi,Arnold:2009ay}.
\subsection{Tree-level FCNC\lowercase{s}}
\begin{figure}
\begin{center}
\begin{fmffile}{tree} 
  \begin{fmfgraph*}(50,20)
  \fmfleftn{i}{2} \fmfrightn{o}{2}
   \fmflabel{$\psi^l$}{i1}
   \fmflabel{$\psi^i$}{i2}
    \fmflabel{$\psi^k$}{o1}
    \fmflabel{$\psi^j$}{o2}
    \fmf{fermion}{v1,i1}
    \fmf{fermion}{v1,o1}
    \fmf{dashes_arrow,label=$\phi$,tension=0}{v1,v2}
    \fmf{fermion}{i2,v2}
    \fmf{fermion}{o2,v2}
     \fmflabel{$\lambda^*_{kl}$}{v1}
      \fmflabel{$\lambda_{ij}$}{v2}
     \fmfdot{v1,v2}
 \end{fmfgraph*}
 \end{fmffile} 
 \end{center}
 \caption{Tree-level exchange of a diquark contributing to $\Delta F = 2$ FCNCs.\label{mix}}
\end{figure}
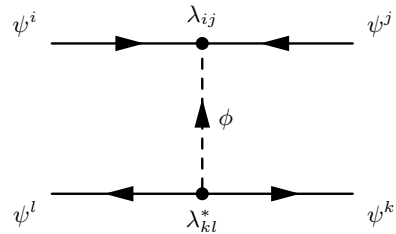

Starting from these eight diquark states, we can reduce the number that may be compatible with our rules of the game to five (with one possible exception), by consideration of tree-level flavour physics processes. 
Indeed, even though all diquark states are charged, their couplings to quarks may allow them to mediate $\Delta F=2$ mixing between neutral mesons at tree-level,
as illustrated in Fig.~\ref{mix}. For example, a canonically-normalized, colour-triplet, electroweak-singlet diquark, $\phi$, of mass $M$, coupled to quarks $\psi_R \in \{u_R, d_R\}$, has the Yukawa interaction
\begin{gather} \label{diqyuk}
\mathcal{L} \supset - \frac{\lambda^\psi_{ij}}{2} \epsilon^{abc} \phi_a \psi^{iT}_{Rb} C \psi^j_{Rc} + \mathrm{h.c.},
\end{gather}
where $a$,$b$, and $c$ are colour indices, and $i $ and $j$ are flavour indices. This generates the dimension-six operator
\begin{gather}  \label{dsix}
\mathcal{L}_{\rm eff} \supset \frac{\lambda^\psi_{ij} \lambda^{\psi *}_{kl}}{4M^2} \overline{\psi^{la}_R} \gamma^\mu \psi^j_{Ra}  \overline{\psi^{kb}_R} \gamma_\mu \psi^i_{Rb},
\end{gather} 
where we have used the antisymmetry of the coupling $\lambda^\psi_{ij}$.

Similarly, for a colour-sextet, electroweak-singlet diquark, $\Phi$, we use a matrix notation in colour space, writing
\begin{gather}
 \Phi =  \begin{pmatrix}\Phi_1          &       \frac{\Phi_4}{\sqrt{2}}     &   \frac{\Phi_5}{\sqrt{2}}  \\
       \frac{\Phi_4}{\sqrt{2}}     &   \Phi_2           &      \frac{\Phi_6}{\sqrt{2}}  \\
               \frac{\Phi_5}{\sqrt{2}}  &    \frac{\Phi_6}{\sqrt{2}}   &     \Phi_3         \end{pmatrix};
\end{gather}
the Yukawa coupling
\begin{gather}
\mathcal{L} \supset - \frac{1}{\sqrt{2}} \lambda^\psi_{ij}  \psi^{iT}_R \Phi  C \psi^j_R + \mathrm{h.c.}
\end{gather}
then generates, via tree-level exchange of the diquark, the same operator (\ref{dsix}), where now $\lambda^\psi_{ij}$ is symmetric.
For any diquark, tree-level $\Delta F =2$ processes can arise via such diagrams only if the diquark can couple to two quarks of the same generation and charge.
As a result, five di-quark states do not mediate such processes.\footnote{This observation was previously made for state VI in \cite{Dorsner:2009mq} and for states IV and VIII in \cite{Gogoladze:2010xd}.} They include the states VI and VIII, whose couplings are purely antisymmetric in flavour indices in the mass basis. Two more are III and IV, which can couple  both to $q_L q_L$ and to $u_R d_R$, since both couplings only connect quarks of different charge.
The fifth and final state is II, which couples antisymmetrically to the $SU(2)_L$ triplet combination of $q_L q_L $, in the gauge basis. This has components coupled to $u_L u_L$ or $d_L d_L$, but these couplings retain antisymmetry in flavour indices in the mass basis. It also contains a third component which couples to quarks of different charges (but is no longer antisymmetric in the mass basis).

We now wish to examine whether the three states that mediate tree-level FCNCs are compatible with our flavour paradigm, {\em viz.} a single, sizable coupling, together with a hierarchical structure. Up until now, we have been rather coy in specifying what we mean by a``sizable coupling''. 
Since we are interested in the prospects for flavourful production at hadron colliders, the most appropriate definition of sizable would seem to be: large enough to result in a statistically-significant sample of signal events at the Tevatron or LHC, after cuts and in the presence of backgrounds and finite experimental resolution. Without entering into a detailed discussion of the experimental analysis, which depends on the specific diquark interaction, we shall simply take the sizable coupling to be unfixed, but of order unity. This will enable readers to keep track in a simple way of  the interdependent scaling of the various indirect bounds and direct production cross-sections.
 
For states mediating tree-level FCNCs, the least dangerous possibility would be to start with the extreme case where all
of the diquark couplings vanish in the gauge basis \cite{Agashe:2005hk}, except for a single sizable coupling, $\lambda^u_{33}$. The rotation that is required to go to the mass basis will then generate a diquark coupling between the first and second quark generations. The smallest this can be is in the case of Chiral Hierarchy (\ref{rmixings}), for diquark V,\footnote{This state was discussed previously in \cite{Mohapatra:2007af}.} in which case we estimate the $1-2$ coupling to be $\sim 8\lambda^u_{33} \times 10^{-5}$. Adapting the bounds of \cite{Isidori:2010kg} for $D$-meson mixing, we find that such a diquark would need a mass of at least $200 \lambda^u_{33}$ GeV. Whereas this is certainly within the reach of the LHC, any other possibility would be marginal. For example, if we moved the large coupling to the $23$ entry, the bound on the mass would increase to $2  \lambda^u_{23}$ TeV. 
A bound of 1.5 $\lambda^u_{33}$ TeV is obtained if we keep $\lambda^u_{33}$ as the large coupling, but switch to CKM-like mixing.
We stress that these are only lower bounds, because one could imagine that the original sub-dominant couplings in the gauge basis were larger than the contributions generated unavoidably by the rotation.

In summary, the most plausible possibility arising from states I, V, and VII occurs when state V has a sizable coupling to $t_R t_R$, in the case of Chiral Hierarchy.
\section{Loop-level Flavour-changing processes\label{sec:loop}}
\subsection{Diquarks II, III, and IV \label{sec:diqloop}}
Even if $\Delta F = 2$ mixing is forbidden at tree level, it will arise at loop level, albeit with a suppression factor, as illustrated (at one-loop) in Fig.~\ref{loopmeson}.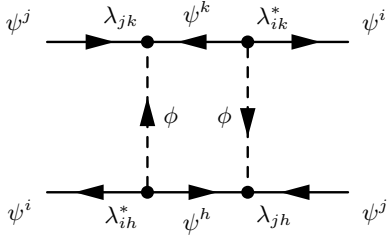
\begin{figure}
\begin{center}
\begin{fmffile}{box} 
  \begin{fmfgraph*}(50,20)
  \fmfleftn{i}{2} \fmfrightn{o}{2}
   \fmflabel{$\psi^i$}{i1}
   \fmflabel{$\psi^j$}{i2}
    \fmflabel{$\psi^j$}{o1}
    \fmflabel{$\psi^i$}{o2}
     \fmflabel{$\lambda^*_{ih}$}{v1}
     \fmflabel{$\lambda_{jh}$}{v3}
     \fmflabel{$\lambda^*_{ik}$}{v4}
      \fmflabel{$\lambda_{jk}$}{v2}
    \fmf{fermion}{v1,i1}
    \fmf{fermion,label=$\psi^h$}{v1,v3}
    \fmf{fermion}{o1,v3}
    \fmf{dashes_arrow,label=$\phi$,tension=0}{v1,v2}
     \fmf{dashes_arrow,label=$\phi$,tension=0}{v4,v3}
    \fmf{fermion}{i2,v2}
    \fmf{fermion,label=$\psi^k$}{v4,v2}
    \fmf{fermion}{v4,o2}
     \fmfdot{v1,v2,v3,v4}
 \end{fmfgraph*}
 \end{fmffile} 
 \end{center}
 \caption{Loop-level exchange of diquarks contributing to neutral meson mixing via chiral operators.\label{loopmeson}}
\end{figure}
Firstly, we note that states III and IV are such that their gauge quantum numbers
allow them to couple to quarks of both chiralities. Unless one of these couplings is somehow suppressed, the $(4 \pi)^2$ suppression of the amplitude relevant for $\Delta m_K$, that comes from the loop factor, will be overwhelmed by the large (factor 400) enhancement coming from the fact that one can now have contributions to non-chiral, $\Delta F=2$ operators (see, for example, Fig.~\ref{nonchibox}), for which the experimental bounds are stronger, due to hadronic and RG effects. (There is, moreover, an enhancement of 1-loop $\Delta F = 1$ processes, such as $b\rightarrow s \gamma$, since the required helicity flip can be placed on an internal top quark.)
\begin{figure}
\begin{center}
\begin{fmffile}{nonchiralbox} 
  \begin{fmfgraph*}(50,20)
  \fmfleftn{i}{2} \fmfrightn{o}{2}
   \fmflabel{$d_{Li}$}{i1}
   \fmflabel{$d_{Rj}$}{i2}
    \fmflabel{$d_{Lj}$}{o1}
    \fmflabel{$d_{Ri}$}{o2}
     \fmflabel{$\lambda^*_{ih}$}{v1}
     \fmflabel{$\lambda_{jh}$}{v3}
     \fmflabel{$\lambda^{\prime*}_{ik}$}{v4}
      \fmflabel{$\lambda^{\prime}_{jk}$}{v2}
       \fmf{fermion}{v1,i1}
    \fmf{fermion,label=$u_{Lh}$}{v1,v3}
    \fmf{fermion}{o1,v3}
    \fmf{dashes_arrow,label=$\phi$,tension=0}{v1,v2}
     \fmf{dashes_arrow,label=$\phi$,tension=0}{v4,v3}
    \fmf{fermion}{i2,v2}
    \fmf{fermion,label=$u_{Rk}$}{v4,v2}
    \fmf{fermion}{v4,o2}
     \fmfdot{v1,v2,v3,v4}
 \end{fmfgraph*}
 \end{fmffile} 
 \end{center}
 \caption{Loop-level exchange of diquarks contributing to neutral meson mixing via non-chiral operators.\label{nonchibox}}
\end{figure}
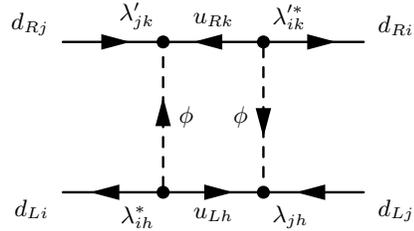
\begin{figure}
\begin{center}
\begin{fmffile}{diwbox} 
  \begin{fmfgraph*}(50,20)
  \fmfleftn{i}{2} \fmfrightn{o}{2}
   \fmflabel{$d_{Li}$}{i1}
   \fmflabel{$d_{Lj}$}{i2}
    \fmflabel{$d_{Li}$}{o1}
    \fmflabel{$d_{Lj}$}{o2}
     \fmflabel{$\lambda^*_{ih}$}{v1}
     \fmflabel{$V^{*}_{hi}$}{v3}
     \fmflabel{$V_{kj}$}{v4}
      \fmflabel{$\lambda_{jk}$}{v2}
    \fmf{fermion}{v1,i1}
    \fmf{fermion,label=$u_{Lh}$}{v1,v3}
    \fmf{fermion}{v3,o1}
    \fmf{dashes_arrow,label=$\phi$,tension=0}{v1,v2}
     \fmf{boson,label=$W$,tension=0}{v4,v3}
    \fmf{fermion}{i2,v2}
    \fmf{fermion,label=$u_{Lk}$}{v4,v2}
    \fmf{fermion}{o2,v4}
     \fmfdot{v1,v2,v3,v4}
 \end{fmfgraph*}
 \end{fmffile} 
 \end{center}
 \caption{One-loop diagram contributing to neutral meson mixing with an internal diquark and $W$-boson.\label{diw}}
\end{figure}
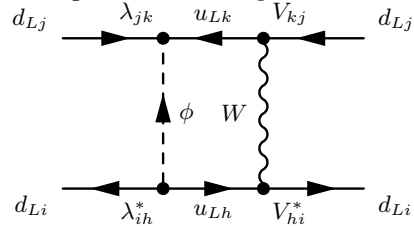
A suppression of one coupling is not unimaginable, however,
in the context of our flavour philosophy: the two couplings involve distinct pairs of SM fermion multiplets ($q_L q_L$ and $u_R d_R$), which may have quite different hierarchies. Moreover, whilst vanishing of one coupling cannot be stable under radiative corrections, a suppression of one coupling relative to the other may be.

So we shall not discard these states just yet, but rather consider the possibility that they couple sizably only to either $q_L q_L$ or to $d_R u_R$. We thus need to consider diagrams of the form in Fig.~\ref{loopmeson}, together with diagrams involving exchange of a diquark and a $W$-boson, as illustrated in Fig.~\ref{diw}. In all cases, we compute contributions to Kaon mixing, where the bound \cite{Isidori:2010kg} is strongest. 

Again, we derive lower bounds by making the extreme assumption \cite{Agashe:2005hk} that all couplings but one vanish in the gauge basis; all other couplings will then be generated by the 
rotation required to go to the mass basis.
If, on the one hand, the dominant coupling is to $q_L q_L$ (states II, III, or IV), then the relevant rotations are expected to be CKM-like. The only safe possibility in this case is to have the large coupling in the $33$ entry, for which the strongest bound comes from the diagram in Fig.~\ref{loopmeson} and yields $M \geq 800 (\lambda^q_{33})^2$ GeV. We note, however, that this is only possible for state IV, since states II and III couple antisymmetrically to $q_L q_L$ in the gauge basis, such that $\lambda^q_{33} \equiv 0$. In contrast, if the large coupling involves only one heavy quark, the same diagram (with a heavy quark in the loop) gives a bound of $M \geq 300 (\lambda^q_{3i})^2$ TeV, where $i \in \{1,2\}$. If the large coupling involves two light quarks, then the same diagram (but with a light quark in the loop) also gives a bound of $300 (\lambda^q_{ij})^2$ TeV, but for two exceptions. The exceptions are states II and III, which couple antisymmetrically in the large diquark mass limit\footnote{As previously mentioned, the rotation to the mass basis maintains the (anti)symmetry in couplings to $u_L u_L$ and $d_L d_L$, but not to $u_L d_L$. Nevertheless, an analogue of the GIM mechanism operates in the diagram in Fig.~\ref{loopmeson} containing a loop of diquarks: only the contribution of the antisymmetric parts of the couplings survives in the limit of large diquark mass.} such that a light quark cannot be exchanged in the loop. In these cases, the strongest bound comes from the diagram in Fig.~\ref{diw}, which yields $M \geq 100 \lambda^q_{ij}$ TeV.

If, on the other hand, the dominant coupling is to $u_R d_R$ (states III and IV), we need to consider separately the cases of CKM-like mixing or Chiral Hierarchy. 
With CKM-like mixing, we derive bounds as above, except that the special cases due to antisymmetric couplings do not arise.
The one safe case, then, is to put the large coupling in the $33$ entry. In the case of Chiral Hierarchy, the necessary rotations always involve the $d_R$ sector, since we start with a single coupling and we wish to change flavour in the $d_R$ sector. Now, the mixings in the $d_R$ sector in the asymmetric case are always comparable to, or greater than, the CKM mixings, as shown in (\ref{rmixings}). 
As a result, the bounds always get tighter. So much so, that even the possibility of putting the large coupling in the $33$ entry becomes unlikely, with a bound of $M \geq 60(\lambda^{ud}_{33})^2$ TeV.

In summary, the only viable possibilities arising from states II, III, and IV, is to couple state IV sizably to $q^3_L q^3_L$, or to couple states III or IV to $t_R b_R$, in the  CKM-like mixing case.
\section{Bounds on antisymmetrically coupled diquarks \label{sec:skew}}
In what follows we concentrate on the remaining states VI and VIII, which are truly antisymmetrically coupled. We do so for the following reasons. 
Firstly, they are the only states that automatically neither mediate tree-level FCNCs, nor contribute to non-chiral operators, nor generate dangerous contributions via one-loop diagrams involving the $W$-boson. Secondly, as we shall see, their antisymmetry allows any one of the three couplings to be the sizable one.
Indeed, as we remarked in the introduction, antisymmetry implies that flavour-changing diagrams must involve all three generations and {\em ergo} at least two couplings, one of which is suppressed by our assumption of a hierarchy. Moreover, the size of the suppression is bounded below by the mixings that arise from the rotation to the mass basis, but the relevant rotations are those of the right-handed quarks and we have seen that these may be much smaller than the mixings of the CKM in the case of Chiral Hierarchy.
 Thirdly, since they have only three coupling constants, it is relatively simply to quote exact, general bounds, which may then be compared with any particular flavour model. Fourthly, as we shall later see, they give suppressed contributions to flavour-diagonal but $CP$ violating processes, such as the EDM of the neutron.

We denote the diquarks VI, VIII by $\phi^{u,d}$.
There are just four independent parameters for each diquark at the renormalizable level: the mass, and the three couplings, which we re-write as
\begin{gather} \label{diqyuk2}
\lambda_{ij}^{u,d} \equiv \epsilon_{ijk} \lambda_{k}^{u,d}.
\end{gather}
 To obtain the most transparent bounds, we translate the bounds on dimension six operators (given, for example, in \cite{Davidson:2007si} and \cite{Isidori:2010kg}) into bounds on the diquark couplings and masses, for each of the three diquarks. In the text, we give the general formul{\ae}, valid for any diquark mass. We then collate the bounds in Table \ref{bounds}, with the simplifying assumption that the mass of the diquark is somewhat greater than the mass of the top quark.
\subsection{Tree-level, flavour-changing decays \label{sec:treedecays}}
The special property of the diquarks VI and VIII, of coupling only antisymmetrically in flavour space, automatically forbids any tree-level contribution to $\Delta F =2$ processes. However, the tree-level exchange of these diquarks can mediate flavour-violating effective interactions involving simultaneously all three generations of quarks. In particular, the exchange of  the diquark VIII leads to
\begin{align}
{\cal L}_{\rm eff}= \frac{\lambda^{d*}_1\lambda^d_3}{2M^2}\left( {\bar b}_R\gamma^\mu s_R~{\bar s}_R\gamma_\mu d_R - 
{\bar b}_R\gamma^\mu d_R~{\bar s}_R\gamma_\mu s_R \right) +\nonumber \\
\frac{\lambda^{d*}_2\lambda^d_3}{2 M^2}\left( {\bar b}_R\gamma^\mu d_R~{\bar d}_R\gamma_\mu s_R - 
{\bar b}_R\gamma^\mu s_R~{\bar d}_R\gamma_\mu d_R \right) + {\rm h.c.}.
\label{effop}
\end{align}

These operators have similarities with the contribution from SM penguin diagrams, but involve purely right-handed currents. Although their coefficients are complex, they cannot interfere with SM amplitudes, which involve left-handed currents, and do not lead to new $CP$ violating effects. Nonetheless, they can lead to interesting effects in two-body, charmless, hadronic decays of $B$ mesons. In particular the operators in the first line of (\ref{effop}) contribute to final states with zero charm and strangeness, which arise in the SM only from highly suppressed penguin diagrams. Interesting processes of this kind are $B \to \phi \pi $, whose branching ratio is estimated to be 1--6$\times 10^{-8}$ in the SM, $B \to \phi \rho $, or the OZI-suppressed process $B \to \phi \phi$, whose SM branching ratio is estimated to be 1--30$\times 10^{-9}$. The decay $B \to \phi \pi $ is considered a good probe of new-physics contributions, since long-distance effects from $B\to KK^*$ rescattering into $\phi \pi$ are expected to be small.

The contribution of the operators in (\ref{effop}) to two-body hadronic $B$ decays can be evaluated using QCD factorization~\cite{Beneke:1999br,Beneke:2000ry}. Following the parametrization used in ref.~\cite{BarShalom:2002sv}, we find
\begin{gather}
BR(B^\pm \to \phi \pi^\pm ) = 2 \times 10^{-3} \left| \lambda^{d*}_1\lambda^d_3\right|^2 \left( \frac{\rm TeV}{M}\right)^4,
\end{gather}  
\begin{gather}
BR(B^0 \to \phi \pi^0 ) = 1 \times 10^{-3} \left| \lambda^{d*}_1\lambda^d_3\right|^2 \left( \frac{\rm TeV}{M}\right)^4,
\end{gather}   
 \begin{gather}
BR(B^0 \to \phi \phi ) = 2 \times 10^{-5} \left| \lambda^{d*}_1\lambda^d_3\right|^2 \left( \frac{\rm TeV}{M}\right)^4.
\end{gather}
The 90\% CL experimental limits are $BR(B^\pm \to \phi \pi^\pm )<2.4 \times 10^{-7}$, $BR(B^0 \to \phi \pi^0 )<2.8 \times 10^{-7}$, and $BR(B^0 \to \phi \phi ) <2 \times 10^{-7}$~\cite{Nakamura:2010zzi}. In this way, we obtain the constraint listed in Table~\ref{bounds}.
\subsection{Flavour bounds on the diquark coupled to $[u_R u_R]$}
For the $\phi^u$, the only bound (except from the forward-backward asymmetry of the top quark pair production at the Tevatron, see below) comes from mixing of neutral $D$-mesons. The basic box diagram generates a four-fermion effective operator in the weak scale Lagrangian of the form
\begin{gather} \label{boxamp}
A_{RR} (\overline{u}_R \gamma^\mu c_R) (\overline{u}_R \gamma_\mu c_R),
\end{gather}
where the colour indices are contracted within the parentheses and the coefficient is given by 
\begin{gather} \label{boxamp1}
A_{RR} = - \frac{1}{32 \pi^2} \left( \frac{\lambda^{u}_{1} \lambda^{u*}_{2}}{ M} \right)^2 G \left(\frac{m_t^2}{M^2}\right),
\end{gather}
where 
\begin{gather}
G(x) \equiv \frac{1 -x^2+ 2x \log x }{(1-x)^3}.
\end{gather}
Note that, since the diquark coupling to quarks is  antisymmetric and the external states are $c$ and $u$, only the $t$ quark can propagate in the loop. 
The imaginary part of the operator coefficient is bounded above by $(2.9 \times 10^3 \; \mathrm{TeV})^{-2}$ (obtained by insisting that the new physics amplitude be no larger than 0.6 of the SM amplitude \cite{Isidori:2010kg}); the resulting bounds on the real and imaginary parts of the combinations of masses and couplings appearing in the coefficient
are given, in the large diquark mass limit, in Table~\ref{bounds}.

Now let us ask which of our various flavour structures are compatible with the bounds. In Table~\ref{inorout}, we estimate the bound on the largest coupling, for each of the three hierarchies and for both  CKM-like mixing (\ref{ckmmixings}) and Chiral Hierarchy (\ref{rmixings}). The bounds assume order-one phases in all diquark couplings. The suppressed mixings in the latter case mean that any one of the three hierarchical structures, normal, inverted, or perverted, could be compatible with an order-one diquark coupling for a diquark mass around a TeV. The bound on the inverted hierarchy with the mixing corresponding to Chiral Hierarchy is particularly weak, allowing a coupling of strength $\sim 4\pi$ in tandem with a mass of a few hundred GeV.
\subsection{Flavour bounds on the diquark coupled to  $[d_R d_R]$}
Bounds for the $\phi^d$ diquark from mixing in $K$ and $B_{d,s}$ systems can be derived in an analogous way and are reported in Table \ref{bounds}.

For $\Delta F =1$ processes, there are bounds 
from $b\rightarrow d \gamma ,s \gamma$ and from $\epsilon^\prime/\epsilon$ in neutral Kaon decays, from one-loop contributions to the $sd$ chromomagnetic operator.

The operator corresponding to photon emission is given by
\begin{gather}
e H^{\dagger} A^{ij}_{LR} \overline{d^i_R} i \sigma^{\mu \nu} q^j_L F_{\mu \nu},
\end{gather}
where 
\begin{gather}
A^{ij}_{LR} =   - \frac{\lambda^d_{i} \lambda^{d*}_{j}}{72 \pi^2 M^2}  \frac{m_j}{v} F\left( \frac{m^2}{M^2}\right),
\end{gather}
\begin{gather}
F(x) \equiv \frac{4-9x +5x^3 + 6x(1-2x) \log x}{4(1-x)^4},
\end{gather}
and $m$ denotes the mass of the quark in the loop. The factor of $m_j$ arises because the necessary helicity flip is on the incoming quark.
We note that the new physics contribution has no interference with the SM amplitude, which has the opposite polarization.

The $sd$ chromomagnetic operator is
\begin{gather}
g_s H^{\dagger} A^{ij}_{LR} \overline{d^i_R} i \sigma^{\mu \nu} G_{\mu \nu}q^j_L ,
\end{gather}
where 
\begin{gather}
A^{ij}_{LR} =\frac{\lambda^d_{i}\lambda^{d*}_{j}}{192 \pi^2 M^2} \frac{m_j}{v} H\left( \frac{m^2}{M^2}\right)
\end{gather}
and
\begin{gather}
H(x) \equiv \frac{1+9x-9x^2 -x^3 + 6x (1+x) \log x }{(1-x)^4}. 
\end{gather}

Now for the bounds. For Kaons, $\delta (\frac{\epsilon^\prime}{\epsilon}) \leq 10^{-3}$, implying a bound on the operator coefficients of \cite{Davidson:2007si}
\begin{gather}
\mathrm{Im} \frac{A^{12}_{LR} - A^{21}_{LR}}{y_s} \leq 10^{-4} \; \mathrm{TeV}^{-2},
\end{gather}
where $y_s$ is the strange Yukawa coupling to the Higgs. The bound for large diquark mass is given in Table~\ref{bounds}.

For $b\rightarrow s \gamma$, the operator bound is $|A^{23}_{LR}| \leq 6 \times 10^{-5} \; \mathrm{TeV}^{-2}$ (obtained by insisting that the new physics amplitude not exceed 15 {\em per cent} of that of the SM \cite{Davidson:2007si}); for $b\rightarrow d \gamma$, we derive a bound of $|A^{13}_{LR}| \leq 3 \times 10^{-5} \; \mathrm{TeV}^{-2}$, by insisting that the new physics amplitude not exceed that of the SM.
 In both cases, the bound for large diquark mass may be found in Table~\ref{bounds}.
Again, the bounds on the largest coupling for each of the three hierarchies may be found in Table~\ref{inorout}. One can see that only the inverted hierarchy can be compatible with the bounds and only then in the case of CKM-like mixing. For state VIII, which couples to down quarks, the strongest bound (at $M=1$ TeV) sometimes comes from $B\rightarrow \phi \pi$. Since this bound relies on the assumption of QCD factorization, we have also quoted the next most stringent bound.

It is worth noting that, in the cases of both $\phi^u$ and $\phi^d$, the weakest flavour bounds occur for the inverted hierarchy, in which the large diquark coupling involves the first two generations. This is a seemingly counterintuitive result, as the light quarks are subjected to the strongest experimental flavour constraints. The reason for this result lies in the antisymmetry of the diquark coupling. Since a third-generation quark must be necessarily exchanged inside the loop, the largest price in mixing angles must be paid in the case of an inverted hierarchy.
\begin{center}
\begingroup
\begin{table}
\begin{ruledtabular}
\begin{tabular}{c  c  }
Process & Bound $/ (M /\mathrm{TeV})$ \\
\hline
$\epsilon_K$ & $\sqrt{|\mathrm{Im} (\lambda^{d}_{1} \lambda^{d*}_{2})^2|} \leq 2.8 \times 10^{-3}$\\
 $\Delta m_K$ & $\sqrt{|\mathrm{Re} (\lambda^{d}_{1} \lambda^{d*}_{2})^2|} \leq 4.6 \times 10^{-2}$\\
$D$ mixing & $\sqrt{|\mathrm{Im} (\lambda^{u}_{1} \lambda^{u*}_{2})^2|} \leq 6.1 \times 10^{-3}$\\
 $D$ mixing & $\sqrt{|\mathrm{Re} (\lambda^{u}_{1} \lambda^{u*}_{2})^2|} \leq 1.5 \times 10^{-2}$\\
$B_d$ mixing & $\sqrt{|\mathrm{Im} (\lambda^{d}_{1} \lambda^{d*}_{3})^2|} \leq 2.0 \times 10^{-2}$\\
 $B_d$ mixing &  $\sqrt{|\mathrm{Re} (\lambda^{d}_{1} \lambda^{d*}_{3})^2|} \leq 3.6 \times 10^{-2}$\\
$B_s$ mixing &$\sqrt{|\mathrm{Im} (\lambda^{d}_{2} \lambda^{d*}_{3})^2|} \leq 1.6 \times 10^{-1}$\\
$\epsilon^\prime/\epsilon$ & $\sqrt{|\mathrm{Im} \lambda^d_{1}\lambda^{d*}_{2}|} \leq 0.37$ \\
$b \rightarrow s + \gamma$ & $\sqrt{|\lambda^d_{2}\lambda^{d*}_{3}|} \leq 1.8$ \\
 $b \rightarrow d + \gamma$ & $\sqrt{|\lambda^d_{1}\lambda^{d*}_{3}|} \leq 0.9$ \\
$R_b$ & $|\lambda^d_{1,2}| \leq 24 $ \\
$A_c$ & $|\lambda^u_{3}| \leq 24 $ \\
 $B^\pm \rightarrow \phi \pi^\pm$ & $\sqrt{|\lambda^d_{3} \lambda^d_{1*}|} \leq 0.1$ \\
\end{tabular}
\end{ruledtabular}
\caption{Bounds in units of $M/$TeV on antisymmetrically-coupled diquarks, valid at large diquark mass (see the text for generally-valid bounds). The couplings are defined in eq'ns (\ref{diqyuk},\ref{diqyuk2}).\label{bounds}}
\end{table}
\endgroup
\end{center}
\begin{center}
\begingroup
\begin{table}
\begin{ruledtabular}
\begin{tabular}{ c  c  c}
Hierarchy & CKM-like & Chiral hierarchy  \\
\hline 
 Inverted & $(\lambda^u_{3})^2 \lesssim 10$ ($D$) & $(\lambda^u_{3})^2 \lesssim 90$ ($D$) \\
Normal & $(\lambda^u_{1})^2 \lesssim 0.03$ ($D$) & $(\lambda^u_{1})^2 \lesssim 0.7$ ($D$) \\
 Perverted & $(\lambda^u_{2})^2 \lesssim 0.03$ ($D$) & $(\lambda^u_{2})^2 \lesssim 0.7$ ($D$)\\
 \hline
 \multirow{2}{*}{Inverted} & $(\lambda^d_{3})^2 \lesssim 2$ ($B_d$) & $(\lambda^d_{3})^2 \lesssim 0.06$ ($K$) \\
 & $\lambda^d_{3} \lesssim 1$ ($B\rightarrow \phi \pi$) & $\lambda^d_{3} \lesssim 0.3$ ($B\rightarrow \phi \pi$) \\
 \hline
  \multirow{2}{*}{Normal, Perverted} &  $(\lambda^d_{1,2})^2 \lesssim 0.01$ ($K$) & $(\lambda^d_{1,2})^2 \lesssim 0.01$ ($K$) \\
   & $\lambda^d_{1,2} \lesssim 1$ ($B\rightarrow \phi \pi$) & $\lambda^d_{1,2} \lesssim 0.3$ ($B\rightarrow \phi \pi$) \\
\end{tabular}
\end{ruledtabular}
\caption{Bounds (with the process in parentheses) on the largest diquark coupling in units of $M/$TeV, for each of the three hierarchies, for CKM-like mixing and Chiral Hierarchy. The couplings are defined in eq'ns (\ref{diqyuk},\ref{diqyuk2}).\label{inorout}}
\end{table}
\endgroup
\end{center}
\subsection{Multiple diquarks\label{sec:mult}}
One question we have not yet addressed is whether one can have multiple diquark states (either coupled to up or down quarks or both) without a contradiction with experimental constraints. 
Could we have, for example, one diquark that couples predominantly to the first and second generations and another that couples predominantly to the first and third?
At least for contributions of the type we have discussed, this would appear to pose no problem, provided the diquark mass eigenstates are not strongly mixed.
If they are strongly mixed, then a single diquark mass eigenstate will have two sizable couplings, which immediately poses a problem for flavour physics. If they are not strongly mixed, then, for example, one-loop contributions to $\Delta F = 2$ processes containing one each of the two diquarks in the loop will still be suppressed, since to get a flavour-changing diagram each diquark propagator must begin and end on different vertices. Explicitly, a one-loop contribution to Kaon mixing, for example, requires both the $1-3$ and $2-3$ coupling for each diquark.
\subsection{Two-loop processes}
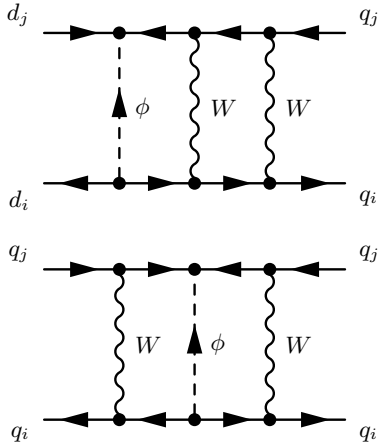
\begin{figure}
\begin{center}
\begin{fmffile}{diw2box} 
  \begin{fmfgraph*}(50,20)
  \fmfleftn{i}{2} \fmfrightn{o}{2}
   \fmflabel{$d_i$}{i1}
   \fmflabel{$d_j$}{i2}
    \fmflabel{$q_i$}{o1}
    \fmflabel{$q_j$}{o2}
    \fmf{fermion}{v1,i1}
    \fmf{fermion}{v1,v3}
    \fmf{fermion}{v3,v5}
    \fmf{fermion}{v5,o1}
    \fmf{dashes_arrow,label=$\phi$,tension=0}{v1,v2}
     \fmf{boson,label=$W$,tension=0}{v3,v4}
      \fmf{boson,label=$W$,tension=0}{v5,v6}
    \fmf{fermion}{i2,v2}
    \fmf{fermion}{v4,v2}
     \fmf{fermion}{v6,v4}
    \fmf{fermion}{o2,v6}
     \fmfdot{v1,v2,v3,v4,v5,v6}
 \end{fmfgraph*}
 \\[3\baselineskip]
   \begin{fmfgraph*}(50,20)
  \fmfleftn{i}{2} \fmfrightn{o}{2}
   \fmflabel{$q_i$}{i1}
   \fmflabel{$q_j$}{i2}
    \fmflabel{$q_i$}{o1}
    \fmflabel{$q_j$}{o2}
        \fmf{fermion}{v1,i1}
    \fmf{fermion}{v3,v1}
    \fmf{fermion}{v3,v5}
    \fmf{fermion}{v5,o1}
    \fmf{dashes_arrow,label=$\phi$,tension=0}{v3,v4}
     \fmf{boson,label=$W$,tension=0}{v1,v2}
      \fmf{boson,label=$W$,tension=0}{v5,v6}
    \fmf{fermion}{i2,v2}
    \fmf{fermion}{v2,v4}
     \fmf{fermion}{v6,v4}
    \fmf{fermion}{o2,v6}
     \fmfdot{v1,v2,v3,v4,v5,v6}
 \end{fmfgraph*}
 \end{fmffile} 
 \end{center}
 \caption{Two-loop diagrams contributing to neutral meson mixing with an internal diquark and $W$-bosons. \label{diw2}}
\end{figure}

For the diquarks coupled to $u_R u_R$ or $d_R d_R$, there are no one-loop contributions involving a diquark and a $W$-boson. Nevertheless, we might worry that there might be strong bounds from two-loop contributions of this type. Such a process need not involve all three generations and so there may be an enhancement that can overcome the extra loop factor.
In Fig.~\ref{diw2}, we show two-loop contributions to Kaon mixing that do not require one to go through a small diquark coupling, in the case where the coupling to the first and second generation quarks is of order one.
It is easy to see that these diagrams give small contributions: the first is a dressing of a GIM-suppressed SM FCNC, together with an insertion of $m_s$ and $m_d$; in the second, the GIM mechanism does not operate, but instead one has four mass insertions, giving a suppression of $(m_s^2 m_d^2)$. 
\section{Other indirect searches \label{sec:diag}}
\subsection{Neutron electric dipole moment}
\begin{figure}
\begin{center}
\begin{fmffile}{diagram1} 
 \begin{fmfgraph*}(50,30)
  \fmfleftn{i}{1} \fmfrightn{o}{1}
   \fmflabel{$d$}{i1}
    \fmflabel{$d$}{o1}
       \fmf{fermion}{i1,v1}
    \fmf{fermion,label=$b,,s$}{v2,v1}
     \fmf{fermion,label=$t$}{v3,v2}
      \fmf{fermion,label=$s,,b$}{v4,v3}
       \fmf{fermion}{v4,o1}
    \fmffreeze
   \fmf{boson,label=$W$,left=1,tension=1}{v2,v3}
    \fmf{dashes_arrow,label=$\phi$,left=0.5,tension=3}{v4,v1}
        \fmfdot{v1,v2,v3,v4}
 \end{fmfgraph*}
 \end{fmffile} 
 \end{center}
 \caption{Two-loop contribution to the neutron EDM.\label{edm}}
\end{figure}
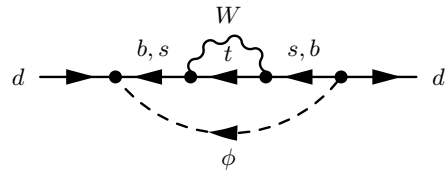

The diquark couplings may contain new sources of $CP$ violation and, {\em ergo}, give new contributions to the electric dipole moment of the neutron. For example, for diquarks coupled antisymmetrically to three generations of quarks, as in the SM, there are potentially three new complex phases in the diquark couplings, with only one new complex degree of freedom (the diquark) that can be re-phased, leading to two $CP$-violating phases. Nevertheless, there are trivially no one-loop contributions, since such diagrams involve the modulus-squared of a {\em single} coupling. At higher-loop level, it is easy to see that, at least for three generations of quarks or fewer, all relevant loop
diagrams containing only quarks and antisymmetrically coupled diquarks can be made real.\footnote{We thank R. Rattazzi for discussions on this point.} Indeed, consider the Lagrangian with three generations of massive quarks and gauge interactions switched off. In the mass basis, there are four possible $U(1)$ phase rotations (one for the diquark and each of the three generations of quarks) that leave all terms in the Lagrangian except the diquark coupling invariant. One linear combination of these corresponds to the conserved baryon number, which also leaves the diquark coupling invariant, but three orthogonal combinations can be used to remove the phases of the three antisymmetric couplings.\footnote{This argument breaks down for $n>3$ generations of quarks, (since then the number of antisymmetric couplings, $\frac{n(n-1)}{2}$, exceeds the number, $n$, of quark generations) or at any $n>1$ for couplings that are not antisymmetric in flavour indices; in both cases one may show that there exist two-loop diagrams with phases that can contribute to the neutron EDM.} With this choice of phases, then, all relevant diagrams are real. Note that, in this basis, the CKM matrix does not take its canonical form and indeed possesses three unremovable phases. Hence, there may be extra sources of $CP$ violation (beyond the SM) once gauge interactions are re-instated. One may show that, even in this basis, there is no dangerous loop contribution from $W$-boson exchange. Potentially, contributions could arise at two-loop level, from diagrams where both a diquark and a $W$ are exchanged, as shown in Fig.~\ref{edm}. However, the contribution of the sum of these diagrams is real. The EDM thus requires three or more loops, along with several quark mass insertions to flip the $d_R$ interacting with the diquark into the $d_L$ interacting with the $W$ and several CKM angle suppressions, and is negligible.

Similar considerations apply to the contribution to the Weinberg operator \cite{Weinberg:1989dx} involving a quark loop, for which one cannot obtain a phase by dressing with diquarks alone. Again, one needs $W$-bosons and Higgs insertions. 
\subsection{Electroweak precision tests}
Bounds on diquarks coupled antisymmetrically to $u_R u_R$ or $d_R d_R$ also arise from one-loop corrections to the coupling of the $Z$ boson to 
charmed or bottom quarks, respectively. The relevant diagrams for the latter have previously been computed in the context of supersymmetric theories with violation of $R$-parity in \cite{Lebedev:1999ze}. At large diquark mass, $M \gg m_Z$, with light quarks in the loop, and assuming a single coupling dominates, we find shifts in the tree-level couplings to the $Z$-boson, $h_{u_R} \equiv -\frac{2}{3} \sin^2 \theta_W, h_{d_R} \equiv \frac{1}{3} \sin^2 \theta_W$, given at leading order by
\begin{align}
\delta h_{d_R} &=  \frac{|\lambda|^2}{72 \pi^2} \sin^2 \theta_W \frac{m_Z^2}{M^2} \log \frac{m_Z^2}{M^2}, \nonumber \\
\delta h_{u_R} &= - 2 \ \delta h_{d_R}.
\end{align}
These shifts in the couplings result in shifts in the measured $R_{b,c}$ and $A_{b,c}$ parameters \cite{:2010vi,Nakamura:2010zzi}, compared to the SM values, given by
\begin{align}
\frac{\delta R_b}{R_b} &\simeq 0.83 \; \delta h_{d_R}, \; \; \frac{\delta A_b}{A_b} \simeq -1.7 \; \delta h_{d_R}, \nonumber \\
\frac{\delta R_c}{R_c}   &\simeq -2.1 \; \delta h_{u_R},  \; \; \frac{\delta A_c}{A_c}   \simeq 5.3 \; \delta h_{u_R}.
\end{align}
The measured values are
\begin{align}
R_b &= 0.21629 \pm 0.00066,  \; \; A_b = 0.923 \pm 0.020,  \nonumber \\
R_c &=  0.1721 \pm 0.0030,  \; \; A_c =  0.670 \pm 0.027.\nonumber \\
\end{align}
Since the SM fit to these observables is rather good, we derive a rough bound by insisting that the diquark contribution not exceed twice the quoted error; the strongest bounds on a TeV mass diquark are listed in Table \ref{bounds}. For the diquark coupled to $u_R u_R$ with a dominant coupling to charm and top quarks, the contribution is increased by roughly $m_t/m_Z$; a precise bound in that case can be obtained using the formul{\ae}  given in \cite{Lebedev:1999ze}.
\section{Direct searches\label{sec:coll}}
To begin the discussion of current and future direct searches for antisymmetrically coupled diquarks, it is useful to consider qualitatively the relative advantages of the different search channels available at $pp$ or $p\overline{p}$ colliders for the different coupling hierarchies. For reference, Table~\ref{xsections14tev} lists leading-order cross sections\footnote{NLO cross-sections may be found in \cite{Han:2009ya}.} for the LHC running at a centre-of-mass energy of 14 or 7 TeV, evaluated for unit diquark coupling, for each of the hierarchies.
\begin{center}
\begingroup
\begin{table*}
\begin{ruledtabular}
\begin{tabular}{c  c  c  c  c  c c  c  c  c  c }
 & \multicolumn{5}{c}{CM Energy =14 TeV} & \multicolumn{5}{c}{CM Energy = 7 TeV}\\
\hline
Mass/GeV & 200 & 400 & 600 & 800 & 1000 & 200 & 400 & 600 & 800 & 1000\\
\hline
 $  gg, q\overline{q} \rightarrow \phi \phi^* $ & $4.6\times 10^1 $ & $1.3\times 10^0 $ & $1.2\times 10^{-1} $ & $2.0\times 10^{-2} $ & $4.4\times 10^{-3}$
 & $7.0\times 10^0 $ & $1.1\times 10^{-1} $ & $6.2\times 10^{-3} $ & $5.6\times 10^{-4} $ & $6.4\times 10^{-5}$ \\
 $ c g \rightarrow \overline{ u } \phi^* $ & $3.5\times 10^2 $ & $2.6\times 10^1 $ & $4.5\times 10^0 $ & $1.1\times 10^0 $ & $3.4\times 10^{-1}$ 
 & $6.9\times 10^1 $ & $3.4\times 10^0 $ & $4.0\times 10^{-1} $ & $7.0\times 10^{-2} $ & $1.6\times 10^{-2}$ \\ 
  $ \overline{ c } g \rightarrow u \phi $ & $3.5\times 10^2 $ & $2.6\times 10^1 $ & $4.5\times 10^0 $ & $1.1\times 10^0 $ & $3.4\times 10^{-1}$
 & $6.8\times 10^1 $ & $3.4\times 10^0 $ & $4.0\times 10^{-1} $ & $7.0\times 10^{-2} $ & $1.6\times 10^{-2}$ \\ 
$ u g \rightarrow \overline{ c } \phi^* $ & $3.6\times 10^3 $ & $4.0\times 10^2 $ & $9.2\times 10^1 $ & $2.9\times 10^1 $ & $1.1\times 10^1$ 
 & $1.1\times 10^3 $ & $9.5\times 10^1 $ & $1.7\times 10^1 $ & $4.3\times 10^0 $ & $1.3\times 10^0$ \\ 

 $ \overline{ u } g \rightarrow c \phi $ & $6.6\times 10^2 $ & $5.3\times 10^1 $ & $9.8\times 10^0 $ & $2.6\times 10^0 $ & $8.5\times 10^{-1}$ 
 & $1.5\times 10^2 $ & $8.1\times 10^0 $ & $1.1\times 10^0 $ & $2.1\times 10^{-1} $ & $5.0\times 10^{-2}$ \\
 $ u c \rightarrow \phi^* \rightarrow u c $ & $1.5\times 10^4 $ & $1.7\times 10^3 $ & $4.2\times 10^2 $ & $1.4\times 10^2 $ & $5.8\times 10^1$ 
 & $6.3\times 10^3 $ & $5.6\times 10^2 $ & $1.1\times 10^2 $ & $3.0\times 10^1 $ & $1.0\times 10^1$ \\ 
 $ \overline{ u} \overline{ c } \rightarrow \phi \rightarrow \overline{ u} \overline{ c } $ & $3.8\times 10^3 $ & $3.1\times 10^2 $ & $5.9\times 10^1 $ & $1.6\times 10^1 $ & $5.7\times 10^0$
 & $1.1\times 10^3 $ & $6.4\times 10^1 $ & $9.2\times 10^0 $ & $2.0\times 10^0 $ & $5.5\times 10^{-1}$ \\  
 $ s g \rightarrow \overline{ d } \phi^*$ & $ 5.0\times 10^2$ & $ 4.0\times 10^1$ & $ 7.1\times 10^0$ & $ 1.9\times 10^0$ & $ 5.9\times 10^{-1}$ 
 & $ 1.1\times 10^2$ & $ 5.8\times 10^0$ & $ 7.4\times 10^{-1}$ & $ 1.4\times 10^{-1}$ & $ 3.2\times 10^{-2}$ \\ 
 $ \overline{ s } g \rightarrow d \phi$ & $ 5.0\times 10^2$ & $ 4.0\times 10^1$ & $ 7.1\times 10^0$ & $ 1.9\times 10^0$ & $ 5.9\times 10^{-1}$ 
 & $ 1.0\times 10^2$ & $ 5.8\times 10^0$ & $ 7.4\times 10^{-1}$ & $ 1.4\times 10^{-1}$ & $ 3.2\times 10^{-2}$ \\
 $ d g \rightarrow \overline{ s } \phi^*$ & $ 2.0\times 10^3$ & $ 2.0\times 10^2$ & $ 4.4\times 10^1$ & $ 1.4\times 10^1$ & $ 5.0\times 10^0$ 
 & $5.7\times 10^2$ & $ 4.4\times 10^1$ & $ 7.3\times 10^0$ & $ 1.7\times 10^0$ & $ 4.9\times 10^{-1}$ \\ 

 $ \overline{ d } g \rightarrow s \phi$ & $ 7.8\times 10^2$ & $6.6\times 10^1$ & $ 1.2\times 10^1$ & $ 3.3\times 10^0$ & $ 1.1\times 10^0$ 
  & $ 1.8\times 10^2$ & $ 1.1\times 10^1$ & $ 1.4\times 10^0$ & $ 2.8\times 10^{-1}$ & $ 6.5\times 10^{-2}$ \\ 
 $ d s \rightarrow \phi^* \rightarrow d s$ & $ 1.1\times 10^4$ & $1.2\times 10^3$ & $ 2.8\times 10^2$ & $ 9.7\times 10^1$ & $ 4.0\times 10^1$ 
 & $4.5\times 10^3$ & $ 3.9\times 10^2$ & $ 7.7\times 10^1$ & $2.1\times 10^1$ & $ 7.2\times 10^0$ \\ 
 $ \overline{ d} \overline{ s } \rightarrow \phi \rightarrow \overline{ d} \overline{ s }$ & $ 5.6\times 10^3$ & $ 5.0\times 10^2$ & $1.0\times 10^2$ & $ 3.1\times 10^1$ & $ 1.1\times 10^1$ & $ 1.9\times 10^3$ & $ 1.3\times 10^2$ & $ 2.0\times 10^1$ & $ 4.7\times 10^0$ & $ 1.4\times 10^0$ \\ 
 $ u g \rightarrow \overline{ t } \phi^* $ & $3.0\times 10^2 $ & $6.0\times 10^1 $ & $1.8\times 10^1 $ & $6.9\times 10^0 $ & $3.0\times 10^0$ 
 & $7.6\times 10^1 $ & $1.2\times 10^1 $ & $2.7\times 10^0 $ & $7.9\times 10^{-1} $ & $2.6\times 10^{-1}$ \\ 

 $ \overline{ u } g \rightarrow t \phi $ & $4.2\times 10^1 $ & $6.6\times 10^0 $ & $1.6\times 10^0 $ & $5.1\times 10^{-1} $ & $1.9\times 10^{-1}$ 
 & $6.9\times 10^0 $ & $7.5\times 10^{-1} $ & $1.3\times 10^{-1} $ & $3.1\times 10^{-2} $ & $8.4\times 10^{-3}$ \\ 

 $ d g \rightarrow \overline{ b } \phi^*$ & $ 5.0\times 10^3$ & $ 4.2\times 10^2$ & $8.5\times 10^1$ & $ 2.5\times 10^1$ & $ 8.9\times 10^0$ 
 & $1.5\times 10^3$ & $ 9.5\times 10^1$ & $ 1.4\times 10^1$ & $3.2\times 10^0$ & $ 8.9\times 10^{-1}$ \\ 

  $ \overline{ d } g \rightarrow b \phi$ & $ 2.1\times 10^3$ & $ 1.4\times 10^2$ & $ 2.4\times 10^1$ & $ 6.1\times 10^0$ & $ 2.0\times 10^0$ 
  & $5.0\times 10^2$ & $2.4\times 10^1$ & $ 2.9\times 10^0$ & $ 5.3\times 10^{-1}$ & $ 1.2 \times 10^{-1}$ \\ 
 $ c g \rightarrow \overline{ t } \phi^* $ & $2.0\times 10^1 $ & $3.0\times 10^0 $ & $7.0\times 10^{-1} $ & $2.1\times 10^{-1} $ & $7.3\times 10^{-2}$ 
 & $2.8\times 10^0 $ & $2.8\times 10^{-1} $ & $4.6\times 10^{-2} $ & $9.7\times 10^{-3} $ & $2.4\times 10^{-3}$ \\ 
 $ \overline{ c } g \rightarrow t \phi $ & $2.0\times 10^1 $ & $3.0\times 10^0 $ & $7.0\times 10^{-1} $ & $2.1\times 10^{-1} $ & $7.3\times 10^{-2}$ 
 & $2.8\times 10^0 $ & $2.8\times 10^{-1} $ & $4.6\times 10^{-2} $ & $9.7\times 10^{-3} $ & $2.4\times 10^{-3}$ \\ 


 $ s g \rightarrow \overline{ b } \phi^*$ & $ 1.3\times 10^3$ & $ 8.6\times 10^1$ & $ 1.4\times 10^1$ & $ 3.4\times 10^0$ & $ 1.1\times 10^0$ 
 & $ 3.0\times 10^2$ & $ 1.3\times 10^1$ & $ 1.5\times 10^0$ & $ 2.7\times 10^{-1}$ & $ 6.1\times 10^{-2}$ \\ 
 $ \overline{ s } g \rightarrow b \phi$ & $ 1.3\times 10^3$ & $ 8.6\times 10^1$ & $ 1.4\times 10^1$ & $ 3.4\times 10^0$ & $ 1.1\times 10^0$ 
 & $ 3.0\times 10^2$ & $ 1.3\times 10^1$ & $ 1.5\times 10^0$ & $ 2.7\times 10^{-1}$ & $ 6.1\times 10^{-2}$ \\ 
\end{tabular}
\end{ruledtabular}
\caption{Leading order cross-sections (in pb) at a 14 or 7 TeV LHC, for unit diquark coupling, computed using {\tt MADGRAPHv5} \cite{mg5,Alwall:2007st} and CTEQ6L1 pdfs \cite{Pumplin:2002vw}. Top and bottom quark pdfs are assumed to vanish.\label{xsections14tev}}
\end{table*}
\endgroup
\end{center} 
\subsection{Inverted hierarchy \label{sec:inv}}
For the inverted hierarchy, in which there is a sizable coupling involving the first and second generation quarks, we have the option of searching either for a resonance in the $s$ channel, or searching for an excess resulting from diquark exchange in the $t$-channel. 
The former channel is initiated by a quark from the first and a quark from the second generation (or a pair of anti-quarks), whereas the latter channel is quark-antiquark initiated, and these may both belong to the first generation.

Thus, for the Tevatron, we might expect 
that the strongest bounds come from $t$ channel exchange searches. Passing to the LHC, the search sensitivity for $t$-channel exchange will be enhanced by the energy reach and luminosity, but will be suppressed by the fact that we need to produce an antiquark out of the sea. CDF, D0, ATLAS and CMS should all be able to present bounds, since they have searched for evidence of the quark contact interaction term, $\mathcal{L} \supset \Sigma_{\mathrm{flavours}} \frac{2 \pi}{\Lambda^2} (\overline{q}_L \gamma^\mu q_L)^2 $ in dijet distributions, quoting lower bounds on $\Lambda$ of 1.6 \cite{Abe:1996mj},  3.1 \cite{:2009mh}, 3.4  \cite{Collaboration:2010eza}, and 5.6 \cite{Khachatryan:2011as} TeV, respectively.\footnote{ATLAS sets a more model-independent limit on the quantity $F_\chi$, which is the ratio of the number of observed events with $\chi \equiv \exp{|y_1-y_2|} < 3.32$ (where $y_{1,2}$ are the jet rapidities) to the number of observed events at any $\chi$, in an event sample with dijet invariant mass exceeding 1.2 TeV. Unfortunately, we expect no more than 0.1 events with such a large dijet invariant mass, for diquark coupling not exceeding one and mass exceeding 200 GeV.} We estimate that the second of these searches translates to bounds of 600 GeV for the diquark coupled to up quarks and 300 GeV for the diquark coupled to down quarks, at unit diquark coupling. We stress however that these estimates are not robust, not least because the assumption of contact interactions is invalidated at such low masses; we regard them as merely suggestive of what might be achieved with a dedicated analysis.
\subsubsection{Search for dijet resonances}
Turning now to dijet resonance searches, these are necessarily quark-quark initiated, but also necessarily require a quark from each of the first and second generations. As a result, there is not necessarily any gain in passing from a $p\overline{p}$ to a $pp$ collider, at least in the limit that the pdfs of strange and charmed quarks equal those of the corresponding antiquarks. Moreover, the enhancement for diquarks coupled to an up quark, rather than a down quark, is mitigated by the fact that one must also couple to a charm quark, which is, of course, somewhat heavier than the strange quark. As a result, we see from Table~\ref{xsections14tev} that cross sections for production of diquark resonances coupled to up-type quarks are similar to those for diquarks coupled to down-type quarks. 

\begin{figure}[t]
  \begin{center} 
    \includegraphics[width=0.99\linewidth]{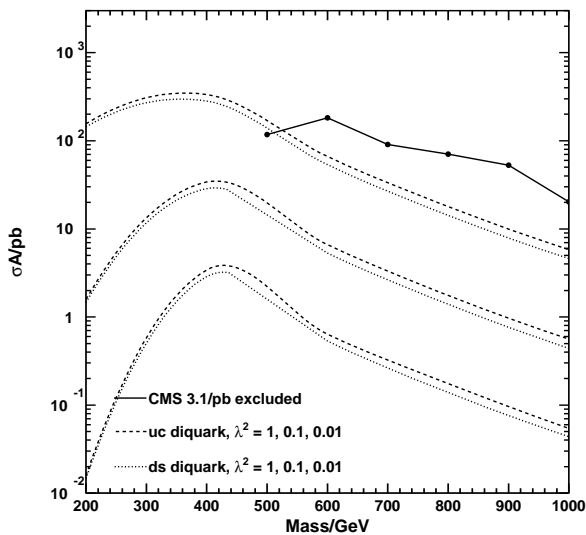}
    \end{center}
\caption{CMS dijet resonance exclusion curve (solid line) \cite{Khachatryan:2010jd} together with leading order predictions for diquarks coupled to $u$ and $c$ quarks (dashed) or to $d$ and $s$ quarks (dotted), with squared coupling of 1, 0.1, and 0.01 (from top to bottom). \label{cmsresonance}} 
\end{figure}

\begin{figure}[t]
  \begin{center} 
    \includegraphics[width=0.99\linewidth]{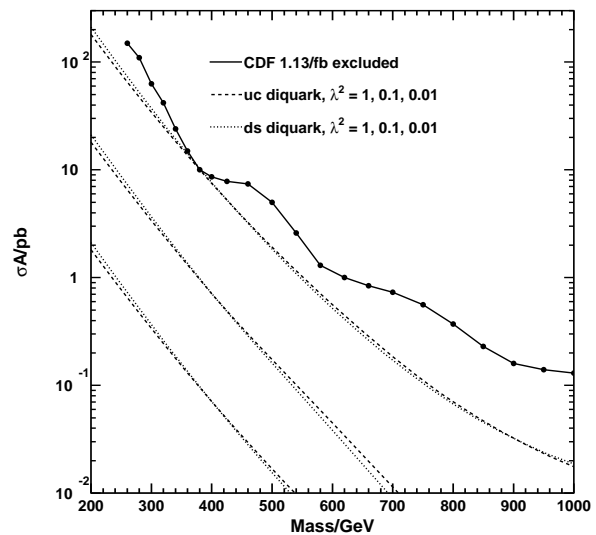}
    \end{center}
\caption{CDF dijet resonance exclusion curve (solid line) \cite{Aaltonen:2008dn} together with leading order predictions for diquarks coupled to $u$ and $c$ quarks (dashed) or to $d$ and $s$ quarks (dotted), with squared coupling of 1, 0.1, and 0.01 (from top to bottom). \label{cdfresonance}} 
\end{figure}

There is, of course, a gain to be had in passing from the Tevatron to the LHC in terms of energy reach.
CMS has performed a search using 3.1 $\mathrm{pb}^{-1}$ of data and presented an exclusion in terms of 
signal cross section times acceptance as a function of resonance mass  \cite{Khachatryan:2010jd}. Since the search is based on `bump-hunting' techniques, 
the sensitivity of the search (and,  {\em ergo}, the bound), can only be independent of the model in the limit that the intrinsic resonance width is somewhat less that the experimental resolution, which is dominated by the jet energy resolution. Moreover, the bound depends on the nature of the initial and final state partons. One effect accounted for in \cite{Khachatryan:2010jd} is that gluon-initiated jets radiate more than quark-initiated ones, leading to a broader bump for resonances involving the former.  CMS estimates that their dijet invariant mass resolution for a quark-quark resonance varies between 5 and 8 {\em per cent} in the region between 0.5 and 1.5 TeV, whereas our diquark width is given (in the limit of decay to massless quarks) by
\begin{gather}
\Gamma = \frac{ |\lambda|^2 }{8 \pi}M.
\end{gather}
For couplings below unity, the width lies below 4 {\em per cent}, below the experimental resolution.\footnote{Table~\ref{inorout} shows that couplings larger than unity are allowed by the flavour bounds; to probe these would require a search for broader resonances.} Assuming implicitly that the search sensitivity is independent of the initial state quark flavours, CMS gives an exclusion curve for a narrow di-quark resonance which we reproduce  (in the region 500 to 1000 GeV) in Fig.~\ref{cmsresonance}. Figure~\ref{cmsresonance} also shows the leading order partonic resonant cross-section\footnote{The $t$-channel exchange diagrams give a small contribution and have been neglected.} times acceptance, for the diquark coupled to up-type or down-type quarks, with a squared-coupling to the first and second generation quarks varying between 0.01 and 1.  We have estimated the signal acceptance by applying all of the cuts described in \cite{Khachatryan:2010jd} directly at the partonic level. Whilst this search, based on a very small data set, only excludes diquarks with order one couplings in a small mass interval for now, it seems that there is hope for the future. 
The lower limit on the search region is set by the trigger, which becomes fully efficient only above 490 GeV. Unfortunately, this lower limit is destined to increase in step with the luminosity.
Turning to the Tevatron Run II, CDF has presented an exclusion based on 1.13 $\mathrm{pb}^{-1}$ of data. The search strategy is similar to the one of CMS, except that CDF uses the quark-gluon resonance profile to set a limit on diquark resonances. Since the former are broader than the latter, due to the different radiation properties of quarks versus gluons, this presumably represents a conservative bound. The CDF exclusion curve, together with the leading order signal cross-section times acceptance (with cuts described in \cite{Aaltonen:2008dn} applied at the partonic level), is shown in Fig.~\ref{cdfresonance}. One sees that no mass region is excluded for couplings smaller than unity on the basis of the leading order signal cross section. 
\subsection{Perverted and normal hierarchies \label{sec:perv}}
\subsubsection{Forward-backward asymmetry in top pair production}
Several authors \cite{Shu:2009xf,Dorsner:2009mq,Dorsner:2010cu,Gresham:2011dg,Patel:2011eh,Grinstein:2011yv,Ligeti:2011vt} have explored the possibility of using diquark exchange in the $t$ channel to explain
the forward-backward asymmetry observed in pair production of top quarks at the Tevatron \cite{Aaltonen:2011kc}.
This requires a rather light diquark (a few hundred GeV), with a coupling to up and top quarks of order a few.

To be compatible with flavour constraints, the other two diquark couplings must be suppressed.\footnote{Some authors \cite{Grinstein:2011yv,Ligeti:2011vt} have extended MFV to diquarks transforming under flavour symmetries (see also \cite{Arnold:2009ay}). This falls outside our flavour paradigm and, as we have seen,
does not seem to be required for consistency with flavour constraints.}  
In the context of our flavour paradigm, Table~\ref{inorout} shows that in the case of Chiral Hierarchy, the tension with the bound from $D$-meson mixing is small. (We stress again that our bounds should only be considered as order of magnitude estimates.) As a result, the diquark explanation seems at least plausible.
For comparison, if we had taken a generic new state mediating $D$-meson mixing at tree level, and coupled non-chirally, with the coupling between the first and second generation quarks generated by the CKM rotation, we would have found a bound on the mass of the new state of $800$ TeV!

We can also use our estimate of the $\lambda^u_{3}$ and $\lambda^u_{1}$ couplings to compare with constraints on single top production.
In \cite{Dorsner:2010cu}, a very rough bound on $|\lambda^u_3 \lambda^u_{1,2}|$ is obtained, by requiring that the LO cross-section for $u \overline{u} \rightarrow \overline{c}t$ 
plus $uc \rightarrow tu$ at the Tevatron not exceed the uncertainty in the measured cross section for single production of top quarks, $\sigma = 2.76^{+0.58}_{-0.47}$ pb \cite{Group:2009qk}, which agrees with the SM prediction for $m_t = 170$ GeV. The bound varies between $|\lambda^u_3 \lambda^{*u}_{2}| < 0.1$ for $M=400$ GeV and $ |\lambda^u_3 \lambda^{*u}_{2}|< 0.9$ for $M = 800$ GeV. Again, there is some tension at lower mass values, but the diquark explanation seems not unreasonable.
\subsubsection{Single {\em vs.} pair production}
For the normal and perverted hierarchies, the (associated) single production cross sections begin to become comparable to the pair production (via QCD) cross sections, such that the latter dominate for small enough diquark couplings and masses. For example, with a unit coupling between the charm and top quarks, Table \ref{xsections14tev} shows that (associated) single production of the diquark dominates for masses above 400 GeV (for both 7 and 14 TeV LHC CM energies), while for a coupling of 0.1, pair production dominates all the way up to a TeV. Alternatively, for a diquark coupled to strange and bottom quarks, single production dominates above 400 GeV at 7 TeV (600 GeV at 14 TeV), even for a coupling as small as 0.1.

In the (associated) single production channel, diquarks present an interesting new signal in the form of a resonance between a heavy-flavour and a light-flavour jet.
As previous authors have discussed, these may be searched for directly \cite{Mohapatra:2007af,Gresham:2011dg}, or, for example, as anomalous excesses in the $t \overline{t} j$ \cite{Mohapatra:2007af,Dorsner:2009mq}
or $b \overline{b} j$ channels. Since the cross-sections can be so large, such new physics may also be able to profit from the use of charm-tagging. 
\subsection{Distinguishing $qq$ from $q\overline{q}$ resonances at the LHC}
The presence of heavy flavour in the final state also allows for the possibility of determining whether a produced resonance coupled to a heavy quark or a heavy anti-quark, for the case of final states involving $b$, $t$, and possibly $c$. With it comes the possibility of distinguishing quark-quark from quark-antiquark resonances.
For example, for a diquark coupled to up quarks with the inverted hierarchy, single production of a diquark resonance at the LHC arises predominantly via a $uc$ initial state (rather than $\overline{u} \overline{c}$) and hence leads to a $uc$ final state, with a predominance of $c$ over $\overline{c}$. But for a neutral quark-antiquark resonance, one would expect equal numbers of $c$ and $\overline{c}$ in the final state, while an electrically-charged quark-antiquark resonance would yield a predominance of $\overline{c}$ in the final state. The differences arise because a diquark or a charged $q\overline{q}$ resonance carries a conserved quantum number (electric charge or baryon number), whereas a neutral $q\overline{q}$ resonance does not.

One may use the same trick for a resonance with the perverted hierarchy produced in association with a heavy quark or antiquark, by focussing on the heavy quark or antiquark in the final state that pairs up with the light quark to form the resonance.

For the normal hierarchy, even in the case of a diquark coupled to $bs$ one may hope to distinguish it from a real quark-antiquark resonance in associated production, notwithstanding the fact that the proton contains comparable fractions of $s$ and $\overline{s}$: in the case of a diquark, the final state will always contain a $b$ and a $\overline{b}$, whilst a real quark-antiquark resonance will lead to equal fractions of same- and opposite-sign $bb$ final states.
\section{Discussion}
The origin of flavour structure remains one of the outstanding puzzles of particle physics.
And yet, 
it seems at first that sub-TeV new physics has  limited prospects for giving us further clues. The 
barrage of flavour-violating and $CP$-violating experimental tests are so constraining 
that the safest course for new physics is
to be predominantly ``flavour-blind", with perhaps small flavour-dependent couplings affecting decays.  
We have argued that new scalars with substantial diquark couplings represent a striking exception, and a special opportunity for the Tevatron and LHC to uncover new flavour physics. Our conclusion is based on a 
judgement on the most plausible and broadly-defined expectation for the flavour dependence of the diquark couplings, namely that they are hierarchical in the same gauge basis as the SM Higgs Yukawa couplings are, but perhaps with a very different hierarchical pattern. Such a structure fits naturally into the dominant UV flavour paradigms. 

We have performed a comprehensive set of estimates based on this structure and demonstrated that several species of diquark can satisfy all flavour and $CP$ constraints, while still having at least one coupling strong enough to play a role in their production. Some of these cases, diquarks with 
$q_L q_L$ or $u_R d_R$ quantum numbers,
can only satisfy the constraints if they are roughly hierarchical in the same way as SM Yukawa couplings, namely with the strongest couplings to the third generation and roughly CKM-like suppressions to other generations.
Colour-triplet diquarks with $u_R u_R$ quantum numbers may exhibit the greatest variety of behaviours, with the strongest couplings between first and third, or first and second, or second and third generations, given the rough flavour structure we denoted as Chiral Hierarchy,  a class of compelling flavour structures
emerging from either Froggatt-Nielsen or extra-dimensional mechanisms. Colour-triplet diquarks with either $u_R u_R$ or $d_R d_R$ quantum numbers can also have their strongest couplings to first and second generations with roughly CKM-like suppressions to the third. 
Curiously, the bounds are weakest when the largest coupling involves the first and second generations, even though the bounds are strongest for processes involving these same quarks externally.  
It is also possible that some of these observations can be extended to some $R$-parity violating supersymmetric theories, but this remains for future work.

The life of a hadron machine, such as the Tevatron or LHC, inevitably involves several tentative signals and anomalies, some of which may emerge as true signals of new physics. While such anomalies certainly need close experimental scrutiny, it is also invaluable to consider at early stages what new physics might underlie them, because it typically indicates where else experiments can look for corroboration. This back-and-forth between experiment and phenomenological modeling has taken place in the course of past anomalies, and should continue going into  the LHC era. Some of the tentative signals have in the past, and will no doubt in the future, suggest new physics coupling directly to quarks. It is at this point that theorists will need to consider the connections between new physics and quarks that can plausibly be in accord with low-energy flavour and $CP$ tests. As we have argued, it is precisely here that diquark scalars come into their own, and the present paper can serve as a valuable resource. 
\section*{Acknowledgments}
We thank K.~Agashe, P.~Agrawal, M.~Baumgart, G.~Buchalla, R.~Frederix, A.~Pomarol, S.~Rahatlou, R.~Rattazzi, and D.~Stolarski for discussions and R.~Contino for participation in the early stages of this work. The research of Raman Sundrum was supported by the United States National Science Foundation 
under grant NSF-PHY-0910467 and by the Maryland Center for Fundamental Physics.

\end{document}